# Chiral metamaterials with PT symmetry and beyond

Sotiris Droulias[1,2*], Ioannis Katsantonis[1,2], Maria Kafesaki[1,2], Costas M. Soukoulis[1,3] and Eleftherios N. Economou[1,4]

[1]*Institute of Electronic Structure and Laser, Foundation for Research and Technology Hellas, 71110 Heraklion, Crete, Greece*
[2]*Department of Materials Science and Technology, University of Crete, 71003 Heraklion, Greece*
[3]*Ames Laboratory and Department of Physics and Astronomy, Iowa State University, Ames, Iowa 50011, USA*
[4]*Department of Physics, University of Crete, 71003 Heraklion, Greece*

Optical systems with gain and loss that respect Parity-Time (*PT*) symmetry can have real eigenvalues despite their non-Hermitian character. Chiral systems impose circularly polarized waves which do not preserve their handedness under the combined space- and time- reversal operations and, as a result, seem to be incompatible with systems possessing *PT* symmetry. Nevertheless, in this work we show that in certain configurations, *PT* symmetric permittivity, permeability and chirality is possible; in addition, real eigenvalues are maintained even if the chirality goes well beyond *PT* symmetry. By obtaining all three constitutive parameters in realistic chiral metamaterials through simulations and retrieval, we show that the chirality can be tailored independently of permittivity and permeability; thus, in such systems, a wide control of new optical properties including advanced polarization control is achieved.

The field of Parity-Time reversal (*PT*) symmetry, initially introduced in the context of quantum mechanics [1-4], has attracted increasing interest because, according to this concept, it is possible even for non-Hermitian Hamiltonians to exhibit real eigenvalues. This occurs if the potential *V* satisfies $V(\mathbf{r}-\mathbf{r}_0) = V^*(-\mathbf{r}+\mathbf{r}_0)$, where **r** is the position operator and $\mathbf{r}_0$ is the point with respect to which the parity transformation is defined (the asterisk denotes the complex conjugate). The eigenvalues remain real below some critical value of the potential, the so-called Exceptional Point, above which they become complex. Because paraxial beam propagation is described by a Schrödinger-like equation, the *PT* concept quickly found fertile ground in optics, where the required loss is naturally occurring and gain can be easily introduced; there, the role of the potential is played by the refractive index *n* and therefore $n(\mathbf{r}-\mathbf{r}_0) = n^*(-\mathbf{r}+\mathbf{r}_0)$ [5-11]. The extension of *PT*-symmetry to systems in which the space reversal is along the propagation direction and the eigenvalues refer to those of the scattering matrix [12-21] led to novel phenomena, such as coherent perfect absorption [13,14], the *PT*-laser absorber [15, 16], and anisotropic transmission resonances [18].

Most recently, some works combined *PT*-symmetry with metamaterials [22-27], which could extend these ideas to new limits, as metamaterials can be designed to have the desired permittivity *ε* and permeability *μ*, at almost any frequency [28]. However, an important class of metamaterials, namely chiral metamaterials, remains highly unexplored in the context of *PT*-symmetry. Chiral metamaterials can be designed to have chirality (*κ*) values orders of magnitude larger than those of natural chiral media [29-36], despite being subwavelength in size; hence, they can replace bulky components in applications related to beam polarization control, such as ultrathin circular polarizers, polarization rotators, wave-plates, polarization-modulators, etc [37].

In this work, prompted by both the enhanced polarization control capabilities of chiral metamaterials and the novel effects associated with *PT*-symmetric systems, we attempt to combine both features in the same system; our goal is to transfer the *PT*-related phenomena to waves of arbitrary polarization, from linear to circular, and to achieve advanced polarization control capabilities, as compared to those offered by passive chiral metamaterials. We find that new optical capabilities are obtained even if the *PT* invariance is relaxed as far as the chirality is concerned. As a first step, we derive the necessary conditions for a chiral system to be *PT*-symmetric; then for the system shown in Fig.1 we obtain the full scattering matrix for arbitrary values of the constitutive coefficients (permittivities, permeabilities, chiralities); next we examine how the obtained general results are modified if *PT* invariance as well as extensions beyond the *PT* invariance are imposed. Finally, by simulations and retrieval we obtain various sets of values of the constitutive coefficients in *realistic chiral metamaterials* possessing either *PT*-symmetry or extensions beyond this symmetry; these sets of values are such as to demonstrate that all our theoretical findings can take place in actual systems.

To find the conditions for *PT*-symmetry in systems with chiral response we cast Maxwell's equations $\nabla \times \mathbf{E} = i\omega\mathbf{B}$, $\nabla \times \mathbf{H} = -i\omega\mathbf{D}$ combined with the proper constitutive equations into an eigenproblem with a Schrödinger-like Hamiltonian (see Supplementary Material for details). The constitutive relations are formulated according to Condon's convention [38] as $\mathbf{D} = \varepsilon\varepsilon_0\mathbf{E}+i(\kappa/c)\mathbf{H}$ and $\mathbf{B} = \mu\mu_0\mathbf{H}-i(\kappa/c)\mathbf{E}$ (*ε*, *μ* refer to the relative permittivity and permeability, respectively, *κ* is the chirality parameter expressing the magneto-electric coupling and *c* the vacuum speed of light). To achieve *PT*-symmetry we require that the Hamiltonian commutes with the *PT* operator and we investigate how circularly polarized waves (the eigenvectors of the corresponding wave equation) are transformed under the action of space- and time- reversal. While time reversal ($t\rightarrow$-*t*) does not affect their handedness, the action of space reversal ($x\rightarrow$-*x*, $y\rightarrow$-*y*, $z\rightarrow$-*z*, by choosing the reflection point at the origin) transforms it between left and right, thus implying an apparent incompatibility of chiral systems with *PT*-symmetry. This limitation can be overcome if systems that respect a reduced spatial parity, such as space reflection by a plane (instead of a point), are considered. Such systems possess material parameters that change for example only along the $z$-direction ($x\rightarrow x$, $y\rightarrow y$, $z\rightarrow -z$); for waves propagating along the $z$-direction, we find that the conditions for the material parameters, so as to have a $PT$-symmetric chiral system, read as follows:

$$\varepsilon(z) = \varepsilon^*(-z), \quad \mu(z) = \mu^*(-z), \quad \kappa(z) = -\kappa^*(-z) \quad (1)$$

The reason for the minus sign in the last relation is the fact that $\kappa$ is a pseudo-scalar since it connects polar vectors (**E** and **D**) to axial vectors (**B** and **H** respectively). The conditions (1) are a generalization of the conditions originally reported in [22] for non-chiral metamaterials and can be realized in practice, e.g. by homogeneously embedding chiral media in loss and gain, as depicted in Fig. 1. The system of Fig. 1 consists of two homogeneous chiral gain/loss slabs which are assumed to be infinite on the $xy$-plane and have finite length along the $z$-direction. Without loss of generality, gain (loss) is assumed to be embedded entirely in the left (right) slab.

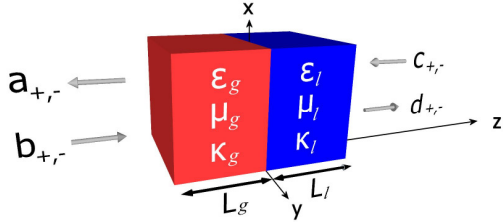

FIG. 1. A system of two homogeneous chiral slabs for the realization of a chiral $PT$-symmetric structure. The complex material parameters $\varepsilon_i$, $\mu_i$ and $\kappa_i$ are the relative permittivity, the relative permeability and the chirality parameter, respectively, and the subscript $i = \{g, l\}$ denotes whether they are located in the '*gain*' or the '*loss*' region. For $PT$-symmetry the shown parameters should satisfy $\varepsilon_g = \varepsilon_l^*$, $\mu_g = \mu_l^*$, $\kappa_g = -\kappa_l^*$ and $L_g = L_l$. The amplitudes of the incident (b,c) and scattered (a,d) waves are shown, where the subscript $+/–$ accounts for Right/Left Circularly Polarized (RCP/LCP) waves.

To study the scattering properties of the double-slab system, we assume that circularly polarized waves arrive at normal incidence from either side of the system (propagating along the $z$-direction) and measure the scattered fields. The amplitudes of the incident (b,c) and scattered (a,d) waves are shown in Fig.1, where the subscript $+/–$ accounts for Right/Left Circularly Polarized (RCP/LCP) waves. Although $PT$-symmetry requirements impose certain relations (see Eq. (1)) between the coefficients in the loss and gain regions, we start with slabs of arbitrary lengths, $L_g$, $L_l$ and arbitrary material parameters $\varepsilon_g$, $\varepsilon_l$, $\mu_g$, $\mu_l$, $\kappa_g$, $\kappa_l$ to obtain general expressions. Due to the two possible circular polarizations at each of the two sides, the system can be described by four input and four output ports and hence by a 4×4 scattering matrix, consisting of eight reflection and eight transmission amplitudes, $\{r, t\}$. Regardless of the side of incidence, we find that $r_{++} = r_{--} = t_{+-} = t_{-+} = 0$, where the second subscript indicates the incident and the first the output polarization. For the rest eight, nonzero scattering coefficients we find:

$$\begin{aligned} r_{L,+-} &= r_{L,-+} \equiv r_L = r_{L,nonchiral} \\ r_{R,+-} &= r_{R,-+} \equiv r_R = r_{R,nonchiral} \\ t_{L,++} &= t_{R,++} \equiv t_{++} = t_{nonchiral} e^{+ik_0(L_g\kappa_g + L_l\kappa_l)} \\ t_{L,--} &= t_{R,--} \equiv t_{--} = t_{nonchiral} e^{-ik_0(L_g\kappa_g + L_l\kappa_l)} \end{aligned} \quad (2)$$

where the subscript $L$ or $R$ denotes incidence from 'Left' or 'Right', respectively, and $k_0$ is the free space wave number (the subscript '*nonchiral*' denotes the respective nonchiral system, i.e. the same system with $\kappa_g = \kappa_l = 0$). From Eqs. (2) it is evident that, although the transmission is the same from both sides, as typically in reciprocal systems [10], the reflection depends on the side of incidence. Note also that, while the reflection amplitudes are always identical to those of the nonchiral counterpart, the transmission amplitudes $t_{++}, t_{--}$ are in general different, but can be made equal to $t_{nonchiral}$ if $L_g\kappa_g + L_l\kappa_l = 0$. In this case the system responds macroscopically as nonchiral, despite having local chirality, i.e. the local chirality is spatially balanced. We emphasize that *the results (2) hold for slabs of general (not necessarily PT) geometrical and material parameters*. The results for the scattering matrix simplify considerably and, as a result, its four eigenvalues appear in two degenerate pairs (see Supplementary Material for details):

$$\lambda_{1,2} = \left(r_L + r_R \pm \sqrt{(r_L - r_R)^2 + 4t_{++}t_{--}}\right)/2 \quad (3)$$

From (2) we notice that $t_{++}t_{--} = t_{nonchiral}^2$ and hence, *the scattering matrix eigenvalues are identical to those of the nonchiral counterpart for slabs of general (non-PT) geometrical and material parameters*.

In general, due to the presence of gain and loss, $|\lambda_{1,2}|\neq 1$. However, if we apply $PT$-conditions an Exceptional Point emerges, below which $|\lambda_1|=|\lambda_2|=1$ ($PT$-phase), while $\lambda_1$, $\lambda_2$ become an inverse conjugate pair above it, satisfying $|\lambda_1||\lambda_2|=1$ [16,18] (broken $PT$-phase). Because the eigenvalues are found to be independent of chirality, the Exceptional Point does not depend on chirality as well and therefore *the PT aspect can be tuned independently from the chiral aspect*. The experimental criterion introduced in [18] for locating the Exceptional Point is now expressed as:

$$\frac{R_L + R_R}{2} - \sqrt{T_{++}T_{--}} = 1 \quad (4)$$

where $R_i \equiv |r_i|^2, i = \{R, L\}$ and $T_j \equiv |t_j|^2, j = \{++,--\}$. In addition, the generalized energy conservation relation [18], which holds both below and above the Exceptional Point, is now expressed as:

$$\left|\sqrt{T_{++}T_{--}} - 1\right| = \sqrt{R_L R_R} \quad (5)$$

According to this expression there exists a flux-conserving scattering process for incident waves on a single side if and only if $\sqrt{T_{++}T_{--}} = 1$, and one of $R_L$ or $R_R$ vanishes, which is known as anisotropic transmission resonance (ATR) [18]. Although this holds for the mean quantity $\sqrt{T_{++}T_{--}}$, this process is not strictly flux-conserving for excitation with only RCP or LCP waves, as $T_{++} \neq T_{--}$ in general. However, inspection of (2) reveals that if $\text{Im}(L_g\kappa_g + L_l\kappa_l) = 0$, i.e. if the imaginary part of $\kappa$ is properly balanced (or zero), then $T_{++} = T_{--} \equiv T$ and flux-conserving ATRs are possible.

As for the chiral features of our system, after applying Eqs. (2) on the optical rotation $\theta = [\arg(t_{++}) - \arg(t_{--})]/2$ and ellipticity $\eta = (1/2)\sin^{-1}[(|t_{++}|^2 - |t_{--}|^2)/(|t_{++}|^2 + |t_{--}|^2)]$ [34], we find that $\theta$ depends only on the real part of $\kappa$, $\text{Re}(\kappa)$, and $\eta$ only on its imaginary part, $\text{Im}(\kappa)$ (see Supplementary Material). For $PT$-symmetric systems and for systems with properly balanced $\text{Re}(\kappa)$ (such as $\text{Re}(L_g\kappa_g + L_l\kappa_l)=0$) $\theta$ is always zero, while the ellipticity of the transmitted wave can be tuned by adjusting $\text{Im}(\kappa)$ in the loss and gain slabs. An alternative particularly interesting case, corresponding to going beyond the full $PT$-symmetry (i.e. only $\varepsilon$ and $\mu$ being $PT$-symmetric), is the case of systems with $\text{Im}(L_g\kappa_g+L_l\kappa_l)=0$. Here $\eta$ is always zero, i.e. the output wave is linearly polarized, while the optical activity can be tuned via $\text{Re}(\kappa)$ (or the length $L$ of the slabs).

To demonstrate the above conclusions, let us consider a system that fulfills conditions (1), such as the system shown in Fig. 1, with $n_g = 2-0.2i$, $n_l = 2+0.2i$ (where $n = \sqrt{\varepsilon\mu}$) and $\kappa_g = -0.165+0.165i$, $\kappa_l = 0.165+0.165i$. The two slabs are assumed to be of equal length $L_g = L_l \equiv L/2$ and the eigenvalues of the scattering matrix, as well as the transmitted and reflected power are shown in Fig. 2a as a function of the normalized frequency $\omega L/c$. The Exceptional Point, which separates the $PT$-phase from the broken $PT$-phase (shaded region), is located at $\omega L/c = 15.4$. Due to the symmetries imposed by Eq. (1) on the chirality parameter $\kappa$, $\text{Re}(\kappa)$ changes sign across the system. Hence, the optical rotation $\theta$ occurring in the first slab is subsequently cancelled when the wave passes through the second slab, resulting overall in zero optical activity, $\theta = 0$, regardless of the location of the Exceptional Point. On the other hand $\text{Im}(\kappa)$ preserves its sign across the entire system and consequently $\text{Im}(L_g\kappa_g + L_l\kappa_l) \neq 0$, resulting in $\eta \neq 0$ and hence $T_{++} \neq T_{--}$, as shown in Figure 2a. Practically, after the ATR located around $\omega L/c = 9.52$ (marked with a vertical dashed line), all higher ATRs are accompanied by $\theta = 0$ and $\eta = 45$ deg. i.e. there exist multiple operation points of zero reflection and full conversion of linearly polarized waves into circularly polarized waves, the handedness of which is controlled by the sign of $\text{Im}(\kappa)$.

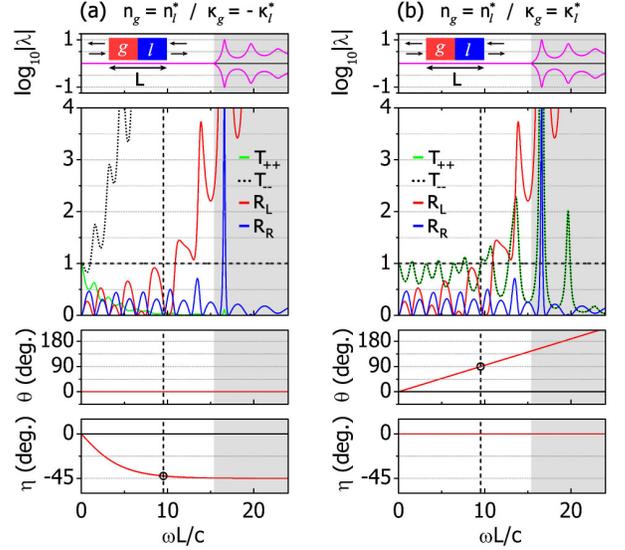

FIG.2. $PT$-chiral system of length $L$ with $n = 2\pm0.2i$. The chirality is (a) $\kappa = \pm0.165+0.165i$ and (b) $\kappa = 0.165\pm0.165i$. Top row: Eigenvalues $\lambda$ of scattering matrix. Middle row: transmittance $T_{++}$, $T_{--}$ and reflectance $R_L$, $R_R$. Bottom rows: optical activity $\theta$ and ellipticity $\eta$. The generalized energy conservation relation (5) is plotted as $R_L R_R + 2\sqrt{T_{++}T_{--}} - T_{++}T_{--} = 1$ (dashed horizontal line). The shaded area denotes the broken $PT$-phase and the dashed vertical line marks an ATR with strong chiral features.

Next, we relax the $PT$-condition (last of Eqs (1)) by setting $\kappa_g = 0.165+0.165i$, $\kappa_l = 0.165-0.165i$; we show below by simulation and retrieval in realistic chiral metamaterials that such choices are actually obtainable and flux-conserving ATRs, as shown in Fig. 2b, are possible. In this case only $\text{Im}(\kappa)$ is spatially odd, leading to $\eta = 0$ (and $\theta \neq 0$) and $T_{++} = T_{--}$. This relaxed condition does not affect the Exceptional Point, as predicted. In this case the ATRs correspond to *unidirectional pure optical rotation of a linearly polarized wave*. At the ATR around $\omega L/c = 9.52$, in particular (marked with a vertical dashed line), $\theta = 90$ deg., i.e. $x$-polarized waves are fully converted to $y$-polarized waves and are transmitted entirely without reflection.

To demonstrate our findings with realistic metamaterials, we employ a chiral metamaterial (CMM) structure similar to the one presented in [35], which consists of two metallic crosses twisted with respect to each other by 30 deg. and embedded in a dielectric host of low index, $n_{host} = 1.41$ (see Fig. 3a, top, for schematic and Supplementary Material for details). To achieve its $PT$ counterpart (referred here as $PT$-CMM), we embed the dielectric host with gain (expressed as the imaginary part of $n_{host}$) and we also tune the relative twist between the two crosses for an additional control on the effective chirality, $\kappa_{eff}$. For $\text{Im}(n_{host}) = -0.09$ and an opposite twist of 30 deg. between the two crosses (see Fig. 3a, bottom), we achieve $PT$-symmetry at 220 THz, as observed in the effective parameters $\varepsilon_{eff}$, $\mu_{eff}$ and $\kappa_{eff}$ of the loss and gain CMMs in Fig. 3b.

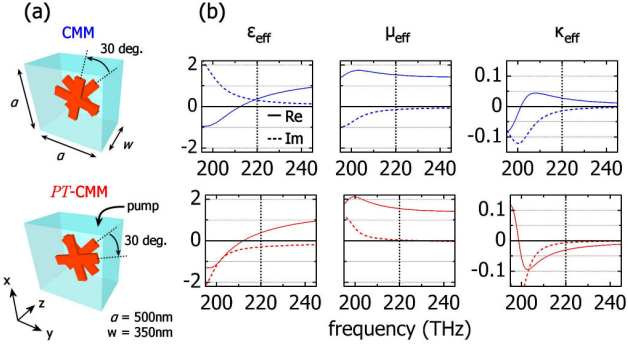

FIG.3. (a) Unit cell of the basic CMM block (top) and its *PT*-counterpart (bottom) satisfying the conditions (1) at 220 THz. The unit cell is periodically repeated on the *xy*-plane. The length of the CMM along *z*-direction is *w*. (b) Retrieved effective permittivity, $\varepsilon_{eff}$, permeability, $\mu_{eff}$, and chirality, $\kappa_{eff}$, as a function of frequency for the structures shown in (a). The results of the top (bottom) row correspond to the CMM (*PT*-CMM) system.

In particular, at this frequency we find $\varepsilon_{CMM} = 0.365+0.299i$, $\varepsilon_{PT\text{-}CMM} = 0.384-0.295i$, $\mu_{CMM} = 1.527-0.149i$, $\mu_{PT\text{-}CMM} = 1.549+0.037i$ and $\kappa_{CMM} = 0.028-0.014i$, $\kappa_{PT\text{-}CMM} = -0.030-0.007i$. To retrieve these parameters we use the procedure outlined in [35], for which we perform full-wave simulations with the commercial software COMSOL Multiphysics (see Supplemental Material for more details).

Next, we keep the *PT*-symmetric CMM-pair intact and we introduce an additional non-magnetic, non-chiral *PT*-slab pair ($\mu = 1$, $\kappa = 0$) of relative permittivity $\varepsilon'\pm i\varepsilon''$, to control the *PT*-potential. The homogeneous slabs have $\varepsilon' = 2.1$ and are of 500nm length each, so that the entire *PT* structure extends over a total of 1.7μm (each CMM is 350nm long). With this configuration, the *PT*-transition can be implemented at a single frequency (220THz) with varying the values of gain and loss, i.e. $\varepsilon''$ (note that the optical potential involves both the frequency and the material parameters). The results are shown in Fig. 4a, where the schematic on top denotes the four different material regions.

As we tune $\varepsilon''$, we observe an Exceptional Point at $\varepsilon''=0.88$ and an ATR close to $\varepsilon''=1$ (with $R_R = 0$). The optical rotation $\theta$ is everywhere close to zero as is imposed by the *PT*-conditions and the available chirality at this frequency leads to a relatively weak ellipticity of $\eta = 2.5$ deg. (this is not a limitation, as $\eta$ can be very large in such systems – see Supplementary Material). The dashed line in the bottom left panel of Fig. 4a is the calculated generalized conservation relation (5). The slight quantitative deviations observed in $\lambda$ and relation (5) from the theoretically expected (see Supplementary Material), come from deviations of the retrieved parameters from the perfect *PT*-conditions. This slight discrepancy is also manifested as a residual optical rotation $\theta$; the observed non-constant ellipticity $\eta$ is due to weak coupling of the CMM blocks with the homogeneous gain/loss slabs, which causes an accordingly weak modification on the effective parameters. However, the positions of the Exceptional Point and the ATR are not affected.

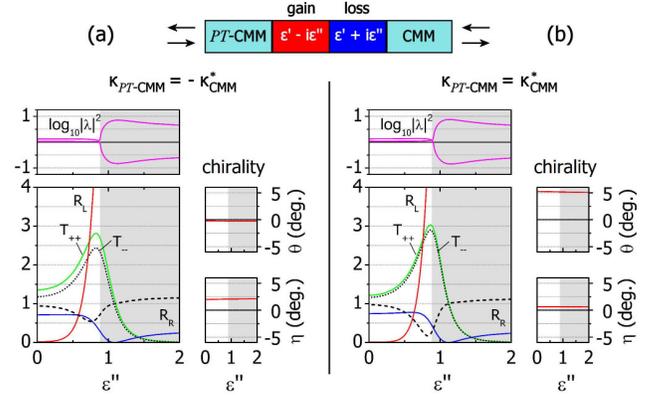

FIG.4. Demonstration of *PT*-transition (Exceptional Point) with a realistic *PT*-chiral metamaterial pair in terms of $\varepsilon''$, the imaginary part of the permittivity of the auxiliary gain/loss slab pair (the real part $\varepsilon' = 2.1$ is constant). The shaded area denotes the broken *PT*-phase. (a) Numerical calculations at 220THz for the CMM and *PT*-CMM blocks of Fig. 3 in the configuration shown in the schematic on top. (b) Results with the same CMM and the modified *PT*-CMM, which has chirality of inverted sign. Panels in each figure: scattering matrix eigenvalues $\lambda$ (top-left), transmittance $T_{++}$, $T_{--}$ and reflectance $R_L$, $R_R$ (bottom-left), optical activity $\theta$ (top-right) and ellipticity $\eta$ (bottom-right). The dashed line in the bottom-left panel is the quantity $R_L R_R + 2\sqrt{T_{++}T_{--}} - T_{++}T_{--}$. The relaxed *PT*-condition in (b) does not affect the Exceptional Point, allowing for the *PT* aspect to be tuned independently from the chiral properties.

Due to $T_{++} \neq T_{--}$, the ATR observed in the system of Fig. 4a is not flux conserving for excitation with only RCP or LCP waves, as mentioned earlier. To achieve the condition $T_{++} = T_{--}$, which corresponds to $\eta = 0$ and thus pure optical rotation $\theta$ for a linearly polarized wave, we need to reverse the sign of chirality in the *PT*-CMM block, so that $\kappa_{PT\text{-}CMM} = (\kappa_{CMM})^*$; this obviously is an extension beyond the *PT* invariance relation (last relation in Eq. (1)). To demonstrate this possibility we reverse the twist between the two crosses in the *PT*-CMM, making it geometrically identical to the basic CMM (see Supplementary Material). This modification changes the sign of $\kappa_{PT\text{-}CMM}$, without practically affecting $\varepsilon_{PT\text{-}CMM}$ and $\mu_{PT\text{-}CMM}$. The new parameters at 220THz are $\varepsilon_{PT\text{-}CMM} = 0.356-0.293i$, $\mu_{PT\text{-}CMM} = 1.538-0.010i$ and $\kappa_{PT\text{-}CMM} = 0.031+0.0071i$. The simulations with the new system are shown in Fig. 4b and verify that the Exceptional Point is not affected by changes in chirality, as predicted by our simple model. Consequently, the properties of *PT*-symmetric systems are still possible, if *PT*-symmetry is obeyed by permittivity or permeability independently of *PT* being satisfied or not satisfied by the chirality, $\kappa$. This conclusion is also verified in the existence of the flux-conserving ATR located close to $\varepsilon''=1$ (with $R_R = 0$).

In conclusion, we have shown that the combination of chirality with *PT* symmetry and even beyond this symmetry is possible in certain systems consisting of actual chiral metamaterials. In such systems the chirality can take a wealth of man-made values even beyond the *PT* symmetry independently of the *PT* conserving values of permittivity and permeability and still maintaining real eigenvalues. Hence, the novel wave propagation and scattering properties of non-chiral *PT*-symmetric systems can be combined with the advanced polarization control properties of chiral metamaterials at will (even if the latter do not obey the *PT* symmetry). Such systems could lead to the realization of novel functionalities, such as coherent laser absorbers for waves of arbitrary polarization and elliptical or circular wave isolators, in a very small scale.

This work was supported by the Hellenic Foundation for Research and Innovation (HFRI) and the General Secretariat for Research and Technology (GSRT), under the HFRI PhD Fellowship grant (GA. no. 4820). It was also supported by the EU-Horizon2020 FET projects Ultrachiral and Visorsurf. Useful discussions with K. Makris are also acknowledged.

*sdroulias@iesl.forth.gr